\documentclass[aps,pra,reprint]{revtex4-1}
\usepackage{amsmath,amsfonts,amssymb}
\usepackage{graphicx,bm}
\usepackage[breaklinks,colorlinks,urlcolor=blue,citecolor=blue,linkcolor=blue]{hyperref}

\newcommand{\vb}[1]{\mathbf{#1}} 

\newcommand{\ket}[1]{| #1 \rangle}

\newcommand{\expval}[1]{\langle #1 \rangle}
\begin{document}
\title{Cavity-induced multispin interactions and phase transitions in ultracold Fermi gases}

\author{Zhen Zheng$^{1,2}$}
\thanks{Corresponding authors: Authors to whom any correspondence should be addressed. Emails: zhenzhen@m.scnu.edu.cn, slzhu@scnu.edu.cn, zwang@hku.hk}
\author{Shi-Liang Zhu$^{1,2,3}$}
\thanks{Corresponding authors: Authors to whom any correspondence should be addressed. Emails: zhenzhen@m.scnu.edu.cn, slzhu@scnu.edu.cn, zwang@hku.hk}
\author{Z. D. Wang$^{1,2,3,4}$}
\thanks{Corresponding authors: Authors to whom any correspondence should be addressed. Emails: zhenzhen@m.scnu.edu.cn, slzhu@scnu.edu.cn, zwang@hku.hk}

\affiliation{
	$^1$Key Laboratory of Atomic and Subatomic Structure and Quantum Control (Ministry of Education), Guangdong Basic Research Center of Excellence for Structure and Fundamental Interactions of Matter, School of Physics, South China Normal University, Guangzhou 510006, China\\
	$^2$Guangdong Provincial Key Laboratory of Quantum Engineering and Quantum Materials, Guangdong-Hong Kong Joint Laboratory of Quantum Matter, Frontier Research Institute for Physics, South China Normal University, Guangzhou 510006, China\\
	$^3$Quantum Science Center of Guangdong-Hong Kong-Macao Greater Bay Area, Shenzhen, China\\
	$^4$HK Institute of Quantum Science \& Technology, The University of Hong Kong, Pokfulam Road, Hong Kong, China
}

\begin{abstract}

The many-body physics of higher-spin systems is expected to host qualitatively new matter phases, but realizing them requires the controllable interactions between multispin components that can be tuned independently for each component.
Here we propose a scheme that meets this demand in ultracold Fermi gases.
By engineering the atom-cavity coupling, we generate cavity-induced effective interactions between pseudo-spin states via multiple Raman and cavity paths.
Focusing on the simplest spin-1 case, we obtain two independent scattering channels whose relative strengths and signs are determined by the Clebsch-Gordan coefficients and optical-field parameters.
The resulting Hamiltonian combines the on-site Cooper pairing with the off-site repulsion,
and drives a continuous transition from the superfluid to the spin-density-wave phase.
The coexistence region is reminiscent of a supersolid, yet the self-organized modulation appears in the spin density profile of a higher-spin representation, rather than in the number density profile.
The proposal can be implemented with the existing techniques in ultracold atoms.
Therefore it offers a versatile platform for quantum simulation of higher-spin many-body physics.

\end{abstract}
\maketitle

\section{Introduction}

The interplay between ultracold atoms and an optical cavity provides an ideal platform for quantum simulating the phase transitions in many-body physics \cite{Bloch2008Jul-rmp,Ritsch2013Apr-rmp,Zhang2018Oct-adv-phys,Mivehvar2021Jan-adv-phys}.
A key to these applications lies in cavity quantum electrodynamics (QED) with atoms, which not only gives rise to spontaneously self-organized orders \cite{Baumann2010Apr-superradiance,Piazza2014Apr-superradiance,Chen2014Apr-superradiance,Zheng2016Oct-superradiance,Colella2018Apr-superradiance,Blass2018Aug-superradiance,Cosme2018Oct-superradiance,Kroeze2018Oct-superradiance,Guo2019May-superradiance,Xu2019May-superradiance,PineiroOrioli2022Mar-superradiance,Zhang2021Aug-superradiance} but also generates photon-mediated effective interactions \cite{Landig2016Apr-cqed-interaction,Guo2012Nov-cqed-interaction,Guo2019May-cqed-interaction,Roux2020Jun-cqed-interaction,Wu2023Dec-cqed-interaction,Luo2024May-cqed-interaction,Paredes2024Sep-cqed-interaction}.
This has motivated a variety of intriguing investigations, including quantum simulation of the Hubbard models \cite{Maschler2005Dec-hubbard,Zhou2011Oct-hubbard,Sheikhan2016Apr-hubbard,Sheikhan2016Dec-hubbard,Zheng2018Feb-hubbard,Colella2018Apr-hubbard,Andolina2019Sep-hubbard}, artificial gauge fields \cite{Kollath2016Feb-gauge-field,Ballantine2017Jan-gauge-field,Mivehvar2017Feb-gauge-field,Fraxanet2023Dec-gauge-field,Bacciconi2024Apr-gauge-field,Dong2014Jan-pra,Deng2014Apr-gauge-field,Zheng2018Feb-njp}, magnetism \cite{Mivehvar2019Mar-magnetism,Colella2019Apr-magnetism}, fermionic superfluids \cite{Schlawin2019Apr-superfluid,Sheikhan2019May-superfluid,Schlawin2019Sep-superfluid,Zheng2020Feb-superfluid,Lewis-Swan2021Apr-superfluid,Hachmann2021Jul-superfluid,Camacho-Guardian2017Nov-superfluid}, and supersolidity \cite{Leonard2017Mar-supersolid,Mivehvar2018Mar-supersolid,Sharma2022Sep-supersolid,Zhang2013Feb-supersolid}.
A significant body of research has focused on the spin-1/2 models, offering direct simulations of electronic systems in solids.
However, since pseudo-spins are typically encoded in the internal states of atoms, engineering models with higher spins is readily achievable in ultracold atomic systems.
This opens up promising avenues for exploring the rich and largely uncharted physics of higher-spin models.

In ultracold-atom systems, Feshbach resonances are routinely exploited to tune the two-body interactions with exquisite precision \cite{Kohler2006Dec-Feshbach,Chin2010Apr-Feshbach}.
Generalization of this control to higher-spin models is conventionally attempted with alkaline-earth(-like) atoms \cite{Gorshkov2010Apr-nphys,Zhang2014Aug-science,Scazza2014Oct-nphys,Sonderhouse2020Dec-nphys,Ono2021Apr-pra,Abeln2021Mar-pra,Tusi2022Oct-nphys,Taie2022Nov-nphys,Pasqualetti2024Feb-prl}, whose metastable excited states supply the additional pseudo-spin levels \cite{Cazalilla2014Nov-rpp,Ludlow2015Jun-alkaline-earth,Ibarra-Garcia-Padilla2024Dec-jpcm}.
Yet the interactions between multispin components in these species arise from the van-der-Waals forces.
It is rather difficult to be adjusted independently for each spin component and thereby necessitates more elaborate external control \cite{Pagano2015Dec-prl,Hofer2015Dec-prl,Yang2022Jun-prr,Yang2022Jun-CommunTheorPhys,Perlin2022Feb-pra,Mamaev2022Aug-prx-quantum,Mukherjee2025Jan-njp}.
On the other hand, cavity QED offers an alternative route.
The photon-mediated interactions emerge from the coherent atom-cavity coupling \cite{Landig2016Apr-cqed-interaction},
and both the scattering amplitude and its sign can be engineered by artificially manipulating the cavity field and the atomic internal states.
This flexibility motivates us to pursue a cavity-QED platform for realizing the tunable interactions between multispin components, circumventing the limitations inherent to the Feshbach-based schemes.

Here, we present a cavity-QED scheme that synthesizes tunable interactions between multispin components in ultracold Fermi gases.
By tailoring the atom-photon coupling, we propose to engineer the effective interactions between pseudo-spin states.
Focusing on a three-component model, we identify two distinct scattering channels whose relative strengths and signs can be tuned by the experimental knobs.
This yields an attraction together with an off-site repulsion, driving a transition from a superfluid to a spin-density-wave (SDW) ordered phase that realizes a higher-spin representation.
The coexisting superfluid and SDW phase is reminiscent of a supersolid \cite{Leonard2017Mar-supersolid}, yet the self-organized modulation appears in the spin density profile rather than in the number density profile.
This proposal is simple and reliable, and its implementations can be realized via current techniques of ultracold atoms.
Therefore, it can offer a practical route to explore and detect the many-body physics of the higher-spin systems.

The paper is organized as follows.
In Sec.\ref{sec:model} we introduce the model Hamiltonian for multispin systems,
and show how the atom-cavity coupling generates an effective interaction that, for the spin-1 case, splits into two independent scattering channels.
These channels allow the coexistence of superfluid and SDW orders, whose phase diagram is analyzed in Sec.\ref{sec:phase}.
In Sec.\ref{sec:exp}, we discuss the realistic implementations of our proposal using ultracold atoms.
Several practical issues regarding the proposal are discussed in Sec.\ref{sec:dis}.
Finally, the brief summary is provided in Sec.\ref{sec:con}.

\section{Model Hamiltonian.} \label{sec:model}

\begin{figure}[t]
	\centering
	\includegraphics[width=0.45\textwidth]{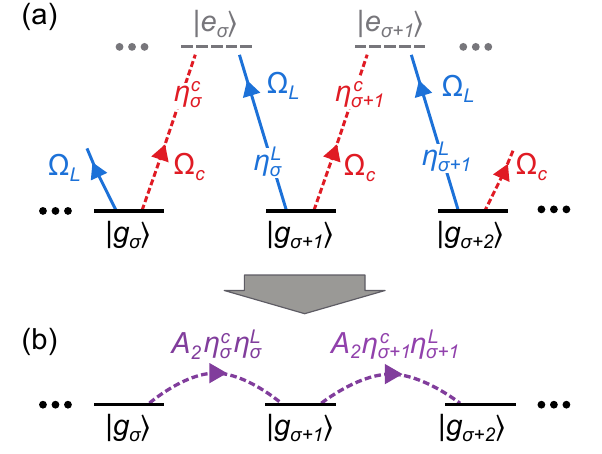}
	\caption{(a) Illustration of the $\Lambda$-type transitions between different spins of $\ket{g}$ through the intermediate state $\ket{e}$.
	The $\pi$-process (red-dashed arrows) is induced by the cavity field $\Omega_c$,
	and the $\sigma^-$-process (blue-solid arrows) is by the laser field $\Omega_L$.
	(b) The mapped model after eliminating $\ket{e}$. Adjacent spins of $\ket{g}$ are coupled with the strength dependent on $\eta_{\sigma}^c$ and $\eta_{\sigma}^L$.}
	\label{fig:Lambda-transition}
\end{figure}

We begin with ultracold Fermi gases that possess multiple internal atomic states, referred as pseudo-spins (or spins for brevity).
These internal states are divided into two groups,
the ground manifold $\ket{g}$ with lower energies and the excited manifold $\ket{e}$ with higher energies.
Atoms in the states $\ket{g_{\sigma}}$ and $\ket{g_{\sigma+1}}$ are coupled through the intermediate state $\ket{e_\sigma}$, as depicted in Fig.\ref{fig:Lambda-transition}(a).
Here $\sigma$ denotes pseudo-spin indexed by the number $\sigma=1,2,3,\cdots,N_s$,
and we assume that the $g$ and $e$ manifolds of pseudo-spins both include the same total spin number $N_s$.
In this $\Lambda$-type transition, we prepare the photon polarization to drive the $\pi$-process via the cavity field and the $\sigma^-$-process via the laser field.
The Hamiltonian of system is then written as
\begin{equation}
	H_{\rm model} = H_{\rm ca} + H_{\rm la} + H_{\rm detuning} \,. \label{eq:hamiltonian-start}
\end{equation}
The first part $H_{\rm ca}$ describes the cavity-induced $\pi$-process \cite{Maschler2008Mar-epjd},
\begin{equation}
	H_{\rm ca} = \sum_{\sigma}\eta_\sigma^c \Omega_{c}a e_\sigma^\dag \psi_{\sigma} + H.c. \label{eq:hamiltonian-start-ca}
\end{equation}
Here $a$ and $a^\dag$ denote the annihilation and creation operators of the cavity photons.
$\psi_{\sigma}$ and $e_{\sigma}$ denote the atomic operators of $\ket{g_\sigma}$ and $\ket{e_\sigma}$.
$\Omega_{c}$ is the strength of the cavity-induced coupling.
Since we focus on the multispin case,
the transition strength is not only determined by the dipole interaction in real space but also evaluated by the transition matrix element in the spin space, i.e., the Clebsch-Gordan (CG) coefficients, which is denoted as $\eta_\sigma^c$.
$H.c.$ stands for the Hermitian conjugate.
The second part $H_{\rm la}$ describes the laser-induced $\sigma^-$-process,
\begin{equation}
	H_{\rm la} = \sum_{\sigma}\eta_\sigma^L \Omega_{L} e_\sigma^\dag \psi_{\sigma+1} + H.c. \label{eq:hamiltonian-start-la}
\end{equation}
where $\Omega_{L}$ is the strength of the laser field, and $\eta_\sigma^L$ stands for the corresponding CG coefficient.
The last part describes the detuning of the above two processes,
\begin{equation}
	H_{\rm detuning} = \Delta_c a^\dag a + \sum_{\sigma}\Delta_e e_{\sigma}^\dag e_{\sigma} \,,\label{eq:hamiltonian-start-detuning}
\end{equation}
where $\Delta_{c}$ and $\Delta_e$ denote the value of the cavity and laser-induced detunings, respectively.

By choosing $\Psi=(\Psi_g,\Psi_e)^T$ with $\Psi_g = (\psi_{1,2,\cdots,N_s})$
and $\Psi_e = (e_{1,2,\cdots,N_s})$, we can cast Hamiltonian (\ref{eq:hamiltonian-start}) into the matrix form,
\begin{equation*}
	H = \Delta_c a^\dag a + \Psi^\dag \begin{pmatrix}
		0 & \hat{M}^\dag \\ \hat{M} & \Delta_e
	\end{pmatrix} \Psi \,,
\end{equation*}
where the block-off-diagonal term $\hat{M}$ is
\begin{equation*}
	\hat{M} = \begin{pmatrix}
		\hat{M}_1 \\ &\hat{M}_2 \\ &&\ddots
	\end{pmatrix} \,,\,
	\hat{M}_\sigma = \begin{pmatrix}
		\eta_\sigma^c \Omega_c a & \eta_\sigma^L \Omega_L \\
		0 & 0
	\end{pmatrix} \,.
\end{equation*}

It may be assumed that the atoms are initially loaded in the states of the $\ket{g}$ manifold and the $\Lambda$-type transition is far detuned.
Thereby the states of the $\ket{e}$ manifold can be adiabatically eliminated.
It gives the effective Hamiltonian expressed in terms of the $\psi$ and $a$ \cite{Zheng2024Sep-multispin-soc}, which reads
\begin{equation}
	H_{\rm model} = \Delta_c a^\dag a - \frac{1}{\Delta_e}\Psi_g^\dag \hat{M}^\dag \hat{M} \Psi_g
	\approx \Delta_c a^\dag a + H_1 +H_2 \,.\label{eq:h-site-space-before-ae}
\end{equation}
Here $H_1$ describes the Stark shift of the spin-$\sigma$ atoms,
\begin{equation}
	H_1 = - \sum_{\sigma} (A_1|\eta_\sigma^L|^2)\psi_{\sigma}^\dag\psi_{\sigma}
\end{equation}
with $A_1=|\Omega_L|^2/\Delta_e$. $H_2$ describes the cavity-induced coupling,
\begin{equation}
	H_2 = - \sum_\sigma (A_2\eta_\sigma^c\eta_\sigma^L) a \psi_{\sigma+1}^\dag\psi_{\sigma} +H.c.\label{eq:h-2-site-space-before-ae}
\end{equation}
with $A_2=\Omega_L^*\Omega_c/\Delta_e$.
One can see that in $H_2$, each spin-$\sigma$ is coupled to the spin-$(\sigma+ 1)$ states, as shown in Fig.\ref{fig:Lambda-transition}(b).
The sign of coupling strength directly depends on $\eta_\sigma^c\eta_\sigma^L$,
which will play a key role in manipulating the cavity-induced interaction.
We remark that in obtaining $H_1$ we have discarded the cavity-photon-induced shift because of the far detuning condition $\Delta_c \gg |\Omega_c|^2/\Delta_e$.

The cavity-induced coupling gives rise to the scattering interaction after further adiabatically eliminating the photon operator $a$.
In this way, we obtain the final form of the effective Hamiltonian as $H_{\rm model}=H_1+H_{\rm int}$,
which is associated with the effective interactions between multispin components,
\begin{equation}
	H_{\rm int} = \sum_{\sigma,\sigma'} U^{\sigma+1,\sigma'}_{\sigma,\sigma'+1} \psi_{\sigma+1}^\dag \psi_{\sigma'}^\dag \psi_{\sigma'+1} \psi_{\sigma} \,. \label{eq:h-int}
\end{equation}
The strength of the cavity-induced interaction depends on the spins of the scattering.
Its form is given as
\begin{equation}
	U^{\sigma+1,\sigma'}_{\sigma,\sigma'+1} = U_{\rm bare} \eta_\sigma^c\eta_\sigma^L \eta_{\sigma'}^c\eta_{\sigma'}^L \,. \label{eq:u-def}
\end{equation}
with the bare strength $U_{\rm bare}=-2|\Omega_L\Omega_c|^2/(\Delta_e^2\Delta_c)$.

\begin{figure}[t]
	\centering
	\includegraphics[width=0.48\textwidth]{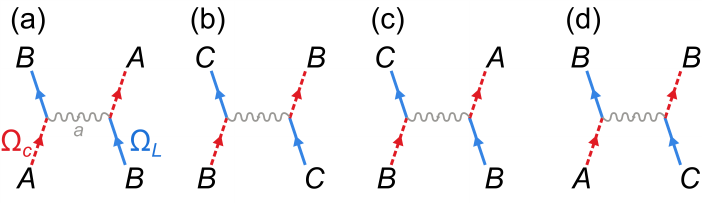}
	\caption{Scattering channels. (a)-(b) The channel originates from the pairwise scattering process.
	(c)-(d) The channel originates from the scattering process that involves three spin components.
	The red-dashed (blue-solid) arrows correspond to the $\pi$ ($\sigma^-$) process in Fig.\ref{fig:Lambda-transition}(a).}
	\label{fig:interaction}
\end{figure}

To clearly convey the essence of our proposal for engineering the multispin scattering, we focus on a model comprising three spin components $\sigma=1,2,3$ hereafter respectively labeled as $\sigma=A,B,C$.
The interacting Hamiltonian (\ref{eq:h-int}) then gives rise to four distinct scattering channels, which are illustrated in Fig. \ref{fig:interaction}. These channels can be categorized into two groups.
(i) The first group, which includes the channels depicted in Fig. \ref{fig:interaction}(a) and (b), describes the conventional scattering between two components. These processes are typical of systems with pairwise interactions.
(ii) By contrast, the second group, which includes the channels shown in Fig. \ref{fig:interaction}(c) and (d), describes the two-body scattering processes involving three distinct components (not genuine three-body interactions).
This group is a crucial distinction in a three-component system compared to a two-component system. The scattering channels within three components introduce additional complexity.
It motivates us to investigate the unique phase transition that are not present in systems with only pairwise interactions.

\section{Phase transitions.} \label{sec:phase}
When the atoms are loaded in a three-dimensional (3D) optical lattice, the system can be described by the tight-binding model,
\begin{equation}
	H = H_0 + H_{\rm int} \,. \label{eq:h-in-lattice}
\end{equation}
Here the first part
\begin{equation}
	H_0 = \sum_{\vb{j},\nu,\sigma} (-t\psi_{\vb{j}+\vb{e}_\nu,\sigma}^\dag\psi_{\vb{j}\sigma} + H.c.) - \sum_{j,\sigma}\mu \psi_{j\sigma}^\dag\psi_{j\sigma}
	\label{eq:h0-in-lattice}
\end{equation}
describes the nearest-neighbor hopping with strength $t$
as well as the chemical potential $\mu$.
$\vb{j}=(j_x,j_y,j_z)$ stands for the vectorized index for the $j$-th site, and $\vb{e}_\nu$ represents the unit vector along the $\nu=x,y,z$ axis.
Hereafter we choose $t$ as the energy unit.
The second part $H_{\rm int}$ is the interacting Hamiltonian from Eq.(\ref{eq:h-int}).
We cast it into the tight-binding representation: $H_{\rm int} = H_{\rm int}^{(1)} + H_{\rm int}^{(2)}$.
Here $H_{\rm int}^{(1)}$ is originated from the conventional pairwise scattering,
\begin{equation}
	H_{\rm int}^{(1)} = \sum_{j,j'} U_0(
		\psi_{j,A}^\dag \psi_{j',B}^\dag \psi_{j',B} \psi_{j,A}
		+\psi_{j,B}^\dag \psi_{j',C}^\dag \psi_{j',C} \psi_{j,B}
	) \,. \label{eq:h-int1-in-lattice}
\end{equation}
Here we have assumed $U_{AB}^{BA}=U_{BC}^{CB}\equiv U_0$,
since the slight mismatches in the interaction strengths could introduce quantitative corrections, but do not alter the underlying qualitative physics picture we aim to describe.
Moreoever, we assume the spatial modulation length of the cavity field exceeds the lattice periodicity, thus the engineered interaction strength can be approximately regarded as uniform.
$H_{\rm int}^{(2)}$ is the characteristic scattering within three components of our model,
\begin{equation}
	H_{\rm int}^{(2)} = \sum_{j}\sum_{j'\neq j}
	U_{1} \psi_{j,A}^\dag \psi_{j',C}^\dag \psi_{j',B} \psi_{j,B} +H.c. \label{eq:h-int2-in-lattice}
\end{equation}
where we also have assumed $U^{CA}_{BB}=U^{BB}_{AC}\equiv U_1$ for simplifying the discussions.
We note that, due to the Pauli exclusion principle, this scattering channel predominantly occurs between two different sites.

It is well known that in Fermi gases, attractive interactions can induce Cooper pairing between atoms of different spin components, leading to a superfluid phase.
As a first step toward a qualitative understanding of interacting Fermi gases, we adopt the Bogoliubov-de Gennes (BdG) mean-field approach.
While this approach neglects thermodynamic fluctuations, it remains qualitatively reliable for 3D fermionic systems and provides the simplest framework for describing the superfluid phase transition.
Notably, one can find that the strength sign in Eq.(\ref{eq:u-def}) is determined by $\Delta_c$, $\eta_\sigma^c$ and $\eta_\sigma^L$.
It paves the way for simultaneously engineering the interactions between multispin components with opposite sign in a single system.
According to Eq.(\ref{eq:u-def}), in the pairwise scattering channel, $U_0$ reduces to the form of $U_0=U_{\rm bare}|\eta_\sigma^c\eta_\sigma^L|^2$,
yielding that the sign of $U_0$ is fixed solely by $U_{\rm bare}$.
By contrast in the scattering channel within three components, $U_1$ reduces to $U_1=U_{\rm bare}\eta_\sigma^c\eta_\sigma^L\eta_{\sigma\pm 1}^c\eta_{\sigma\pm 1}^L$,
hence the sign of $U_1$ depends not only on $U_{\rm bare}$ but also on the CG coefficients,
offering an additional lever for artificial manipulations.
For example, by preparing opposite signs between $\eta^c_\sigma$ and $\eta^c_{\sigma+1}$ involved in Fig.\ref{fig:Lambda-transition},
the interaction $U_{0}$ in Eq.(\ref{eq:h-int1-in-lattice}) is attractive ($U_0<0$ if $U_{\rm bare}<0$), while the off-site interaction $U_{1}$ in Eq.(\ref{eq:h-int2-in-lattice}) can be engineered to be repulsive ($U_1>0$).
These interactions not only sustain the superfluid order but can additionally induce the density-wave order, giving rise to supersolidity, the phase characterized by spontaneous incommensurate density modulations relative to the underlying lattice potential.

Based on the above analysis, we introduce different order parameters to cast the interacting hamiltonian into quadratic forms.
For the superfluid order parameters, it is known that the off-site Cooper pairing induces a $k$-dependent correction, corresponding to hybrid Cooper pairing with $s$-, $p$-, and $d$-wave characteristics.
Nevertheless, in spin-balanced systems, the $s$-wave pairing remains dominant under local attractive perturbations \cite{Schlawin2019Sep-superfluid} and
hereafter we simply consider the on-site superfluid order parameters derived from Eq.(\ref{eq:h-int1-in-lattice}),
\begin{equation}
	\expval{\psi_{j,B}\psi_{j,A}}\equiv \Delta_1/U_0 \,,\, \expval{\psi_{j,C}\psi_{j,B}}\equiv \Delta_2/U_0 \,. \label{eq:order-parameter-superfluid}
\end{equation}
For the Hamiltonian (\ref{eq:h-int2-in-lattice}), the interaction must be off-site, and the order parameters is introduced as
\begin{equation}
	\expval{\psi_{j,A}^\dag \psi_{j,B}} \equiv m_1+(-1)^{j}\delta_1 \,,\,
	\expval{\psi_{j,B}^\dag \psi_{j,C}} \equiv m_2+(-1)^{j}\delta_2 \,.\label{eq:order-parameter-spin}
\end{equation}
To capture the essential physics of the order parameter (\ref{eq:order-parameter-spin}),
we select the basis $\Psi_{j}=(\psi_{j,A},\psi_{j,B},\psi_{j,C})^T$, and find that the spin polarization on the $j$-th site is calculated as
\begin{align*}
	\expval{\hat{S}_x}_j &= \expval{\Psi_{j}^\dag \hat{S}_x \Psi_{j}} \qquad (\hat{S}_x = \frac{1}{\sqrt{2}}\begin{pmatrix}
		0 & 1 & 0 \\ 1 & 0 & 1 \\ 0 & 1 & 0
	\end{pmatrix})\\
	&=\sqrt{2} \cdot{\rm Re}[m_1+m_2
	+ (-1)^j (\delta_1+\delta_2)] \,,\\
	\expval{\hat{S}_y}_j &= \expval{\Psi_{j}^\dag \hat{S}_y \Psi_{j}} \qquad (\hat{S}_y = \frac{1}{\sqrt{2}}\begin{pmatrix}
		0 & -i & 0 \\ i & 0 & -i \\ 0 & i & 0
	\end{pmatrix})\\
	&=\sqrt{2} \cdot{\rm Im}[m_1+m_2
	+ (-1)^j (\delta_1+\delta_2)] \,.
\end{align*}
It reveals that the order parameters $m_{1,2}$ and $\delta_{1,2}$ indeed characterize the density of spins on each site.
Specifically, the first term $m_{1,2}$ represents the uniform spin polarization in the $x$-$y$ plane of the spin space.
For our three-component system, $m_{1,2}$ captures the net magnetization of the spin-1 model.
The second term $\delta_{1,2}$ indicates the modulations of the spin polarizations, correspondingly.
Here we consider the simplest case, i.e. the spatially alternating pattern.
The formula $(-1)^{j}$ in Eq.(\ref{eq:order-parameter-spin}) equals to $(-1)^{j_x}\cdot (-1)^{j_y}\cdot (-1)^{j_z}$ with $j_{\nu=x,y,z}$ denoting the site index projected along the $\nu$ direction.
When $\delta_{1,2}\neq 0$, it gives rise to SDW,
a periodic modulation of spin density that breaks translational symmetry in the real space.
We note that whereas the spin-1/2 model is known to exhibit the SDW order \cite{Masalaeva2021Feb-spin-half-sdw},
here we treat the higher-spin generalization.
Moreover, although the definitions of Eqs.(\ref{eq:order-parameter-superfluid}) and (\ref{eq:order-parameter-spin}) is based on the approximation that is the globally uniform, allowing local order parameters only produces spatial modulations near the lattice boundaries without modifying the bulk phase diagram at the mean-field level.

After introducing the order parameters (\ref{eq:order-parameter-superfluid}) and (\ref{eq:order-parameter-spin}),
Hamiltonian (\ref{eq:h-in-lattice}) can be cast into a quadratic form.
Taking the presence of $\delta$ into considerations,
the momentum $\vb{k}$ of atoms will be transferred to $\vb{k}+\vb{K}$,
where $\vb{K}=(\pi,\pi,\pi)/d$ if we set $\hbar=1$ and $d$ represents the lattice constant.
Consequently, to derive the BdG Hamiltonian within a complete basis framework, we select the basis $\Phi_{\vb{k}}=(\Psi_{\vb{k}},\Psi_{\vb{k}+\vb{K}},\Psi_{-\vb{k}},\Psi_{-\vb{k}-\vb{K}})^T$
with $\Psi_{\vb{k}}=(\psi_{\vb{k},A},\psi_{\vb{k},B},\psi_{\vb{k},C})$.
In the momentum-$\vb{k}$ space, the BdG Hamiltonian derived from Eq.(\ref{eq:h-in-lattice}) is given as (see Appendix \ref{sec:app:bdg})
\begin{equation}
	H_{\rm BdG}(\vb{k}) = \begin{pmatrix}
		\hat{D} & 2I_2\otimes\hat{\Delta} \\
		2I_2\otimes\hat{\Delta} & -\hat{D}
	\end{pmatrix} \,. \label{eq:h-bdg}
\end{equation}
Here $\hat{D}=\xi_{\vb{k}}I_2\otimes I_3 + 2U_{1}L_c(I_2\otimes\hat{m}-2(\sigma_+\otimes\hat{\delta}+H.c.))$
with $\xi_{\vb{k}}=-2t\sum_{\nu=x,y,z}\cos(k_\nu d)-\mu$ and $L_c$ denoting the interaction-range cutoff (see Appendix \ref{sec:app:bdg}).
$\sigma_{x,y,z}$ are Pauli matrices and $I_N$ denotes the $N\!\times\!N$ identity matrix.
$\otimes$ stands for the Kronecker product.
The forms of other matrices are
\begin{equation*}
	\hat{m} = \begin{pmatrix}
		0 & m_2 & 0 \\
		m_2 & 0 & m_1 \\
		0 & m_1 & 0
	\end{pmatrix} ,\,
	\hat{\delta} = \begin{pmatrix}
		0 & \delta_2 & 0 \\
		0 & 0 & 0 \\
		0 & \delta_1 & 0
	\end{pmatrix} ,\,
	\hat{\Delta} = \begin{pmatrix}
		0 & \Delta_1 & 0 \\ 0 & 0 & \Delta_2 \\ 0 & 0 & 0
	\end{pmatrix} .
\end{equation*}

The ground state of the system is determined by the thermodynamic potential $\Omega$, whose form is given in Appendix \ref{sec:app:bdg}.
By self-consistently minimizing $\Omega$ with respect to $\Delta_{1,2}$, $m_{1,2}$ and $\delta_{1,2}$,
we can determine the ground state of the system and obtain the phase diagrams.

\begin{figure}[t]
	\centering
	\includegraphics[width=0.48\textwidth]{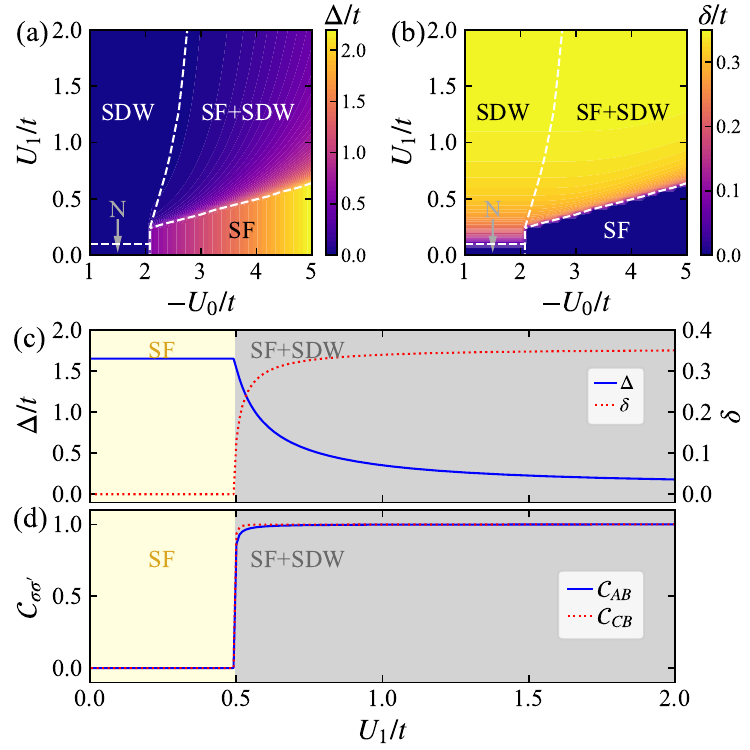}
	\caption{(a) The order parameters $\Delta$ and (b) $\delta$ as functions of $U_0$ and $U_{1}$.
	The colors characterize the magnitudes of $\Delta$ and $\delta$.
	SF stands for the superfluid phase, and N stands for the normal gas phase with $\Delta=\delta=0$.
	The white-dashed lines indicate the phase boundaries.
	We set $\mu=1.2t$.
	(c) The order parameters $\Delta$ and $\delta$, and (d) the correlation function $\mathcal{C}_{AB}$ and $\mathcal{C}_{CB}$ as functions of $U_{1}$ at $U_0 = -4.0t$, corresponding to panels (a)-(b). 
	In (d), $\mathcal{C}_{AB}$ and $\mathcal{C}_{CB}$ are normalized to their maximum absolute values.}
	\label{fig:order-parameters}
\end{figure}

We remark that while slight mismatches in the interaction strengths between different scattering channels could introduce quantitative corrections, they do not alter the underlying qualitative physics picture we aim to describe.
This is because these order parameters are directly governed by the interactions and such mismatches in the interaction strengths will merely introduce the spin-dependent order parameters (i.e. $\Delta_{1}\neq\Delta_{2}$).
Therefore to simplify the investigations, we have assumed that the interaction $U_0$ is spin-independent in Eq.(\ref{eq:h-int1-in-lattice}), and thus the order parameters reduce to $\Delta_{1,2}\equiv\Delta$, $m_{1,2}\equiv m$ and $\delta_{1,2}\equiv \delta$.
In Fig.\ref{fig:order-parameters}(a)-(b), the phase diagram in the $U_0$-$U_{1}$ plane is presented.
We observe that the order parameter $\Delta$ grows monotonically with $U_0$, whereas $\delta$ rises monotonically with $U_{1}$ once $U_{1}$ exceeds a threshold (see Fig.\ref{fig:order-parameters}(c)).
By contrast, our calculation finds that $m$ is identically zero across the entire parameter plane, signalling the absence of the net spin polarization.
This reflects our choice to restrict the analysis to the simplest case with $U^{BA}_{AB}=U^{CB}_{BC}$,
and relaxing this constraint would generally yield $m\neq 0$ and SDW is still present.
In the regime of strong $U_0$ and weak $U_{1}$, the system is in the conventional superfluid phase.
As $U_{1}$ increases, the system undergoes the transition to a phase where superfluid and SDW orders coexist as shown in Fig.\ref{fig:order-parameters}(c).
This suggests that in this phase region, although the pairing is uniform in real space, the spin polarization should break the translational symmetry in real space.
We remark that the coexistence of the superfluid and SDW orders is the reminiscent of the supersolidity,
but the spontaneous modulation is self-organized in the spin density profile rather than the number density profile.
The SDW order persists even in the weak $U_0$ regime (see Fig.\ref{fig:order-parameters}(b)).
In this phase, the system exhibits antiferromagnetic properties of the spin-1 model.

We remark that the mechanism of the spin-supersolid in this work differs from the usual spinor condensate \cite{Kawaguchi2012Nov-spinor-bec}.
In a spinor condensate, the superfluid and magnetic orders originate from the same condensate wavefunction. In the coexistence region, the superfluid and SDW orders involve different scattering channels and are stabilized by distinct cavity-induced interactions. They coexist but are not simply two facets of a single order parameter.
Furthermore, the mechanism of the spin-supersolid via the cavity-mediated channel-selective interactions also differs from the usual supersolid mechanisms, which is originated from the dipolar interactions \cite{Bruun2008Dec-prl}.

Due to the spatially modulated nature of the SDW order parameter, the phase transition can be detected through the correlation function $\expval{n_{\vb{k}\sigma}n_{\vb{k}'\sigma'}}$ with $n_{\vb{k}\sigma}=\psi_{\vb{k}\sigma}^\dag \psi_{\vb{k}\sigma}$ denoting the density operator.
In practice, it can be extracted by making the Fourier transformation to the density correlation function $\expval{n_{\sigma}(\vb{r})n_{\sigma'}(\vb{r}')}$ that can be measured in a time-of-flight technique.
Specifically, at the mean-field level, we calculate the quantity $\mathcal{C}_{\sigma\sigma'}
= \expval{n_{\vb{k}\sigma}n_{\vb{k}'\sigma'}} - \expval{n_{\vb{k}\sigma}}\expval{n_{\vb{k}'\sigma'}}
\approx - \expval{\psi_{\vb{k}\sigma}^\dag \psi_{\vb{k}'\sigma'}}\expval{\psi_{\vb{k}'\sigma'}^\dag \psi_{\vb{k}\sigma}}$ with $\vb{k}'=\vb{k}+\vb{K}$ \cite{Block2014Oct-supersolid}.
As shown in Fig.\ref{fig:order-parameters}(d), we can find that the correlation functions $\mathcal{C}_{AB}$ and $\mathcal{C}_{CB}$ is nonzero whenever  $\delta\neq 0$, making them suitable indicators for identifying the phase transition.

\section{Experimental Realization.} \label{sec:exp}
\begin{figure}[t]
	\centering
	\includegraphics[width=0.48\textwidth]{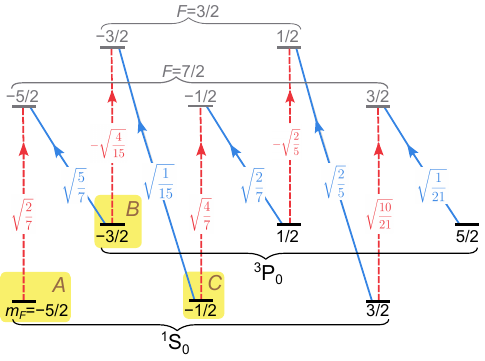}
	\caption{Illustration of the setups for $^{173}$Yb.
	The red-dashed (blue-solid) arrows correspond to the $\pi$ ($\sigma^-$) process in Fig.\ref{fig:Lambda-transition}.
	The number beside each arrow indicates the CG coefficients during the transitions.}
	\label{fig:experimental-setup}
\end{figure}

We now delve into the experimental realization of engineering the interactions between multispin components.
Here, we use the alkaline-earth-like atoms $^{173}$Yb as an example to present our proposal.
It has the perfect decoupling of the nuclear spin $I=5/2$ from the electronic angular momentum $J=0$, which gives rise to six internal states (i.e. the total angular momentum $F=5/2$) for both its ground term $^1$S$_0$ and the excited term $^3$P$_0$, as shown in Fig.\ref{fig:experimental-setup}.
For the alkaline-earth-like atoms, $^3$P$_0$ provides metastable internal states with a long lifetime on the order of milliseconds, and the magic wavelength lattice ensures that $^1$S$_0$ and $^3$P$_0$ states are loaded in the same lattice \cite{Ludlow2015Jun-rmp}.
We alternatively use the hyperfine states of $^1$S$_0$ and $^3$P$_0$ to represent $\ket{g_\sigma}$ in the model shown in Fig.\ref{fig:Lambda-transition},
and respectively employ higher excited states with $F=7/2$ and $3/2$ as the intermediate state $\ket{e_\sigma}$.
Specifically for the three-component model studied in Sec.\ref{sec:phase}, we choose the hyperfine states with $m_F=-5/2$ and $-1/2$ of $^1$S$_0$ as the pseudo-spin $A$ and $C$,
and $m_F=-3/2$ of $^3$P$_0$ as $B$, respectively.
The atoms are initially prepared in these three spin components.
Subsequently, the coupling between spin $A$ (or $C$) and $B$ is generated via the $\Lambda$-type transition through the intermediate excited state $\ket{F,m_F}=\ket{7/2,-5/2}$ (or $\ket{3/2,-3/2}$), respectively.
As illustrated in Fig.\ref{fig:experimental-setup}, it is worth highlighting that the CG coefficient associated with the transition between $^3$P$_0$ and the $F=3/2$ excited states exhibits a negative value, in contrast to the positive values observed for the other transitions depicted.
When preparing $U_{\rm bare}<0$,
it results in $U_{1}=U_{CA}^{BB}=U_{\rm bare}\cdot[\sqrt{\frac{2}{7}}\sqrt{\frac{5}{7}}(-\sqrt{\frac{4}{15}})\sqrt{\frac{1}{15}}]>0$ in Eq.(\ref{eq:h-int2-in-lattice}),
$U_{AB}^{BA}=U_{\rm bare}\cdot(\sqrt{\frac{2}{7}}\sqrt{\frac{5}{7}})^2<0$
and $U_{BC}^{CB}=U_{\rm bare}\cdot(-\sqrt{\frac{4}{15}}\sqrt{\frac{1}{15}})^2<0$
used in Eq.(\ref{eq:h-int1-in-lattice}).
Hence the setups support to simultaneously engineer the attractive and repulsive interactions.

We next assess the experimental feasibility of the parameters employed in the phase diagrams.
To confine $^{173}$Yb atoms, we construct the 3D optical lattice by counter-propagating laser fields with wavelength $\lambda_{\rm OL}=759$nm.
Using the lattice recoil energy $E_R = h^2/(2m\lambda_{\rm OL}^2) \approx 2\pi\hbar\times 2.0\text{kHz} \approx 96\text{nK}$ as our energy unit, we set the lattice depth to $V_L = 3.0E_R \approx 2\pi\hbar\times 6.0$kHz.
This yields a hopping amplitude $t \approx 0.11E_R$ \cite{Walters2013Apr}.
By tuning $(\Delta_e, \Omega_L, \Delta_c, \Omega_c)=2\pi\hbar\times(50,5,500,7)$Mhz,
the bare interaction strength $U_{\rm bare} \approx -2.0E_R$.
Then we obtain $U_{AB}^{BA}\approx -2.4t$, $U_{BC}^{CB}\approx -4.3t$, and $U_{1}\approx 3.2t$.
These values are well within the parameter regime predicted to support the coexistence of superfluid and SDW phases, thus validating our theoretical framework.
We remark that the decay effect in the $\Lambda$-type transitions of Fig.\ref{fig:experimental-setup} can be considered as an additional imaginary part of $U_{\rm bare}$.
However, by preparing the cavity decay rate $\kappa_c=2\pi\hbar\times 1$MHz and the spontaneous emission $\kappa_e=2\pi\hbar\times 100$kHz, these imaginary parts are of the order $10^{-2}$ less to the real part of $U_{\rm bare}$ and hence are ignorable.

\section{Discussions} \label{sec:dis}

We note that although mean-field theory provides a useful guide for capturing the essential physics, the order parameters are indeed affected by fluctuations. For example, in the attractive Hubbard model, beyond-mean-field results \cite{Koga2011Aug-pra} show that superfluid order is suppressed in the presence of fluctuations, yet still survives up to a critical temperature on the order of the hopping magnitude $t$. Furthermore, regarding the SDW order which is manifested in terms of correlated density, the analogous supersolid phase has also been estimated to persist up to 1.6$t$ \cite{Zeng2014May-prb}. 
The critical temperature that supports the phase diagram is thus estimated in order of $10$ nK under the setups of Sec.\ref{sec:exp}.
On the other hand, previous theoretical works \cite{Chen2005Jun-physrep} predict 0.2$T_\text{F}$ is the critical temperature for the fermionic superfluid phase, and the experiment \cite{Wu2023Dec-cqed-interaction} reports that the cooled temperature reaches 0.21$T_\text{F}$ for fermionic $^6$Li atoms. Here $T_\text{F}$ is the Fermi temperature and depends on atom species. For $^{173}$Yb, $T_\text{F}\sim 100$ nK \cite{Taie2010Nov-prl} gives a critical temperature $T_c\approx 0.2T_\text{F}\sim 10$nK, which is consistent with our estimated temperature in order of 10nK and within the range of demonstrated cooling capabilities.
We remark that by keeping the detunings large or the cavity finesse high, it can effectively
suppress heating and reduce the effect of thermal fluctuations to a negligible level,
then the mean-field phase diagram remains essentially valid.

Since the cavity field mode generally exhibits a standing-wave pattern, this induces a spatial modulation of the interactions and affects the specific wavevector of the SDW order. Nevertheless, as the competition between the pairing and density-wave channels persists, we expect that the precise phase boundaries of the coexistence region may shift depending on the specific parameters,
but the superfluid-SDW transition will remain. This expectation is supported by our control calculations with different interaction ranges (Appendix \ref{sec:app:bdg}), where the qualitative phase structure remains unchanged even when the interaction profile is varied.
Furthermore, the methodology developed for the 3D system can be extended to the 2D case. In this scenario, a uniform interaction can be approximately realized by aligning the cavity axis perpendicular to the 2D atomic plane.

Our model is presented in a simplified case where the detuning $\Delta_e$ is the same for all excited states $\ket{e_\sigma}$, as shown in Eq.(\ref{eq:hamiltonian-start-detuning}).
From Eq.(\ref{eq:u-def}), it can be seen that the bare interaction $U_{\rm bare}$ depends on $\Delta_e$.
If $\Delta_e$ is made spin-dependent, this opens up the possibility for independently tuning the effective interaction strengths $U_{AB}^{BA}$, $U_{BC}^{CB}$, $U^{CA}_{BB}$ and $U^{BB}_{AC}$.
In this way, the phase diagram shown in Fig.\ref{fig:order-parameters} will be spin-dependent and  can be systematically explored.

In the whole letter, we restrict the discussion to three pseudo-spin levels to keep the physics transparent.
The scheme can be extended to higher spins by loading atoms into extra pseudo-spins.
From Fig.\ref{fig:experimental-setup}, one can find that adjacent pseudo-spins still couple with alternating signs of CG coefficients for the $\pi$-process.
Hence the pairwise scattering process remains attractive while the scattering process within three components stays repulsive.
The superfluid-SDW phase transition will still exist, albeit with the re-definition of the SDW order parameter corresponding to the enlarged spin representation.
Our approach therefore provides a ready platform for exploring SDW orders in arbitrary-spin representations.

\section{Conclusions} \label{sec:con}
We have proposed a scheme to synthesize effective interactions in systems with more than two pseudo-spin states.
The engineering of the interactions between multispin components relies on the cavity-induced coupling, enabling simultaneous control over the sign of the interaction strength in different scattering channels.
As a result, this approach can be further applied to investigate the coexistence of superfluid and SDW phases.
The scheme is both simple and feasible with current experimental techniques, and thus holds promise for exploring the many-body physics of higher-spin models.

\section*{Acknowledgments}
We thank Yan-Xiong Du, Chang Li, Zhenyu Wang, and Wu Bian for helpful discussions.
This work was supported by National Key Research and Development Program of China (Grant No. 2022YFA1405300),  Innovation Program for Quantum Science and Technology (Grant No. 2021ZD0301700),  Guangdong Provincial Quantum Science Strategic Initiative (Grant Nos. GDZX2304002, GDZX2404001), the NSFC/RGC JRS grant (Grant No. N\_HKU 774/21),  GRF (Grant No. 17303023) of Hong Kong, and the Guangdong Basic and Applied Basic Research Foundation (Grant No. 2026A1515012354).

\appendix

\section{BdG Approach}\label{sec:app:bdg}

In this section, we present the details to the self-consistent solution procedure used in the main texts.
By introducing the order parameters in Eqs.(\ref{eq:order-parameter-superfluid}) and (\ref{eq:order-parameter-spin}),
Hamiltonian (\ref{eq:h-int1-in-lattice}) reduces to the following form,
\begin{align}
	H_{\rm int}^{(1)} &= \sum_j(\Delta_1\psi_{j,A}^\dag\psi_{j,B}^\dag+\Delta_2\psi_{j,B}^\dag\psi_{j,C}^\dag+H.c.) \notag\\
	&- (|\Delta_1|^2+|\Delta_2|^2)/U_{0} \,, \label{eq:app:h-int1-site-space}
\end{align}
and Hamiltonian (\ref{eq:h-int2-in-lattice}) reduces to
\begin{align}
	H_{\rm int}^{(2)} &= \sum_{\vb{j},\nu} \sum_{l=1}^{L_c} U_1(\psi_{\vb{j},A}^\dag\psi_{\vb{j}+l\vb{e}_\nu,C}^\dag \psi_{\vb{j}+l\vb{e}_\nu,B}\psi_{\vb{j},B} \notag\\
	&+\psi_{\vb{j},C}^\dag\psi_{\vb{j}+l\vb{e}_\nu,A}^\dag \psi_{\vb{j}+l\vb{e}_\nu,B}\psi_{\vb{j},B}) + H.c. \notag\\
	&= \sum_{\vb{j},\nu} \sum_{l=1}^{L_c} U_1\Big\{[m_1+(-1)^\vb{j}\delta_1]\psi_{\vb{j}+l\vb{e}_\nu,C}^\dag \psi_{\vb{j}+l\vb{e}_\nu,B} \notag\\
	&+[m_2+(-1)^{\vb{j}+l\vb{e}_\nu}\delta_2]\psi_{\vb{j},A}^\dag \psi_{\vb{j},B} \notag\\
	&+[m_1+(-1)^{\vb{j}+l\vb{e}_\nu}\delta_1]\psi_{\vb{j},C}^\dag \psi_{\vb{j},B} \notag\\
	&+[m_2+(-1)^{\vb{j}}\delta_2]\psi_{\vb{j}+l\vb{e}_\nu,A}^\dag \psi_{\vb{j}+l\vb{e}_\nu,B}+\varepsilon_0 \Big\}+ H.c. \label{eq:app:h-int2-site-space}
\end{align}
Here to handle the cavity-induced long-range interaction \cite{Landig2016Apr-cqed-interaction}, we introduce $L_c$ as an interaction-range cutoff to facilitate the numerical calculations and set $L_c=10$ in Fig.\ref{fig:order-parameters}.
The constant $\varepsilon_0=2(-m_1m_2 + \delta_1\delta_2)$.
After making the Fourier transformation,
they are written as
\begin{align}
	H_{\rm int}^{(1)} &=\sum_\vb{k} (\Delta_1\psi_{\vb{k},A}^\dag\psi_{-\vb{k},B}^\dag+\Delta_2\psi_{\vb{k},B}^\dag\psi_{-\vb{k},C}^\dag+H.c.) \notag\\
	&- (|\Delta_1|^2+|\Delta_2|^2)/U_{0} \,, \label{eq:app:h-int1}
\end{align}
and
\begin{align}
	H_{\rm int}^{(2)} &= \sum_\vb{k} \sum_{l=1}^{L_c}  U_1\Big\{
	2m_1 \psi_{\vb{k},A}^\dag\psi_{\vb{k},B}
	+2m_2 \psi_{\vb{k},C}^\dag\psi_{\vb{k},B} \notag\\
	&+\delta_1 \psi_{\vb{k},A}^\dag(\psi_{\vb{k}+\vb{K},B}+\psi_{\vb{k}+l\vb{K},B})\notag\\
	&+\delta_2 \psi_{\vb{k},C}^\dag(\psi_{\vb{k}+\vb{K},B}+\psi_{\vb{k}+l\vb{K},B})+\varepsilon_0
	\Big\}+ H.c. \label{eq:app:h-int2}
\end{align}
where $\vb{K}=(\pi,\pi,\pi)/d$ if we set $\hbar=1$, $d$ represents the lattice constant,
and the summation $\sum_{\vb{k}}$ is taken over the first Brillouin zone.
Correspondingly by transforming Eq.(\ref{eq:h0-in-lattice}) into the $k$ space, we obtain
\begin{equation}
	H_0(\vb{k}) = \sum_{\sigma=A,B,C}\xi_{\vb{k}}\psi_{\vb{k},\sigma}^\dag\psi_{\vb{k},\sigma} \label{eq:app:h0}
\end{equation}
with $\xi_{\vb{k}}=-2t\sum_{\nu=x,y,z}\cos(k_\nu d)-\mu$.
We select the Nambu basis $\Phi_{\vb{k}}=(\Psi_{\vb{k}},\Psi_{\vb{k}+\vb{K}},\Psi_{-\vb{k}}^\dag,\Psi_{-\vb{k}-\vb{K}}^\dag)^T$
with $\Psi_{\vb{k}}=(\psi_{\vb{k},A},\psi_{\vb{k},B},\psi_{\vb{k},C})$.
Under this basis,
the total Hamiltonian of the model system is obtained by combining Eqs.(\ref{eq:app:h-int1}), (\ref{eq:app:h-int2}), and (\ref{eq:app:h0}), which is written as $H = \sum_{\vb{k}} \Phi_{\vb{k}}^\dag H_{\rm BdG}(\vb{k}) \Phi_{\vb{k}} + \Omega_0$.
Here $H_{\rm BdG}(\vb{k})$ is known as the BdG Hamiltonian whose form has been given by Eq.(\ref{eq:h-bdg}).
The energy constant $\Omega_0= \sum_{\vb{k}}3\xi_{\vb{k}}/4- (|\Delta_1|^2+|\Delta_2|^2)/U_{0}+2U_{1}(-m_1m_2+\delta_1\delta_2 +H.c.)$.
We note that for the expression $3\xi_{\vb{k}}/4$ in $\Omega_0$, the factor 3 arises from summing over the three spins, while the factor 4 arises from both the combination due to Nambu doubling and the pairing between the subscripts $\vb{k}$ and $\vb{k}+\vb{K}$ in the basis $\Phi_{\vb{k}}$.
The thermodynamic potential is calculated as
\begin{equation}
	\Omega = \Omega_0 + \frac{1}{4}\sum_{\vb{k},\alpha} E_\alpha(\vb{k})\Theta(-E_\alpha(\vb{k})) \,,
\end{equation}
where the $E_\alpha(\vb{k})$ is the $\alpha$-th eigen-value of the matrix form $H_{\rm BdG}(\vb{k})$ ($\alpha=1,2,3,\cdots,12$),
and $\Theta(\cdot)$ is the Heaviside-step function that describes the Fermi-Dirac distribution at zero temperature.

To obtain the information of order parameters for the ground state,
we self-consistently minimize the thermodynamic potential $\Omega$.
This is achieved by solving the following equations
\begin{equation}
	\frac{\partial \Omega}{\partial \Delta_{1,2}}
	=\frac{\partial \Omega}{\partial m_{1,2}}
	=\frac{\partial \Omega}{\partial \delta_{1,2}} = 0 \,.
\end{equation}
For the simplified case that sets $\Delta_{1,2}=\Delta$, $m_{1,2}=m$ and $\delta_{1,2}=\delta$,
the above six self-consistency equations can reduce to three, greatly speeding the numerics while retaining the essential physics.

\begin{figure}[t]
	\centering
	\includegraphics[width=0.48\textwidth]{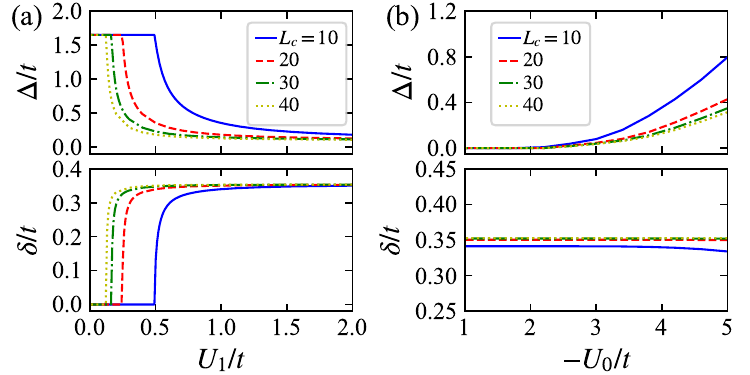}
	\caption{The order parameters $\Delta$ (top) and $\delta$ (bottom) as functions of (a) $U_{1}$ and (b) $U_0$.
	The interaction-range cutoff $L_{c}$ is set as 10 (blue-solid line, i.e. the original data from Fig.\ref{fig:order-parameters}), 20 (red-dashed line), 30 (green-dash-dotted line), and 40 (yellow-dotted line).
	We set $U_0 = -4.0t$ in (a) and $U_1=1.0t$ in (b).}
	\label{fig:order-parameters-longer-range}
\end{figure}

In Sec.\ref{sec:phase}, we have studied interactions with a range cutoff $L_c​=10$ to illustrate the physics of the phase transitions. As the cavity-induced interaction is intrinsically long-ranged \cite{Landig2016Apr-cqed-interaction}, we hereby demonstrate that the essential physics is already captured by this model and remains robust when longer-range interactions are retained. In Fig.\ref{fig:order-parameters-longer-range}, we display the numerical results for various $L_c$ values. We find that the phase transitions reported in Sec.\ref{sec:phase} remain valid: the superfluid and SDW orders still coexist.
For the phase boundary between the superfluid-dominated regimes, it exhibits that the transition point shifts to zero (at a decreasing rate) with the increase of $L_c$, as shown in Fig.\ref{fig:order-parameters-longer-range}(a),
while the phase boundary between the SDW-dominated regimes is insensitive to the choice of $L_c$ as shown in Fig.\ref{fig:order-parameters-longer-range}(b).
This yields that the extent of the pure superfluid phase region shrinks depending on the finite system size,
but the conclusions regarding the coexisting superfluid and SDW phase drawn in Sec.\ref{sec:phase} hold even when longer-range interactions are included.

\section{Effective interactions}\label{sec:app:int}

To obtain the form of the effective interaction (\ref{eq:u-def}), we need to adiabatically eliminate the cavity photon $a$ in Eq.(\ref{eq:h-site-space-before-ae}).
For the Heisenberg motion equation of $a$, we have
\begin{equation}
	i\hbar\partial_t a = [a,H_{\rm model}] = \Delta_c a - \sum_\sigma (A_2\eta_\sigma^c\eta_\sigma^L)^* \psi_{\sigma}^\dag \psi_{\sigma+1} \,.
\end{equation}
By setting $\partial_t a=0$, we obtain $a = \sum_\sigma (A_2^*\eta_\sigma^c\eta_\sigma^L/\Delta_c)\psi_{\sigma}^\dag \psi_{\sigma+1}$ and insert it into Hamiltonian (\ref{eq:h-2-site-space-before-ae}) \cite{Bonifacio2024Dec-prxquant}.
It gives
\begin{align}
	H_2 &= - \sum_{\sigma} \frac{|A_2|^2}{\Delta_c}\eta_\sigma^c\eta_\sigma^L
	\sum_{\sigma'} \eta_{\sigma'}^c\eta_{\sigma'}^L\psi_{\sigma'}^\dag \psi_{\sigma'+1} \psi_{\sigma+1}^\dag\psi_{\sigma} +H.c. \notag\\
	&= - \sum_{\sigma,\sigma'} \frac{2|A_2|^2}{\Delta_c}
	\eta_\sigma^c\eta_\sigma^L \eta_{\sigma'}^c\eta_{\sigma'}^L
	\psi_{\sigma+1}^\dag\psi_{\sigma'}^\dag \psi_{\sigma'+1} \psi_{\sigma} \\
	&\equiv \sum_{\sigma,\sigma'} U_{\rm bare} \eta_\sigma^c\eta_\sigma^L \eta_{\sigma'}^c\eta_{\sigma'}^L
	\psi_{\sigma+1}^\dag\psi_{\sigma'}^\dag \psi_{\sigma'+1} \psi_{\sigma} \label{eq:app:h-2}
\end{align}
where we have denoted
\begin{equation}
	U_{\rm bare} = -\frac{2|A_2|^2}{\Delta_c}=-\frac{2|\Omega_L\Omega_c|^2}{\Delta_e^2\Delta_c} \label{eq:app:u-bare}
\end{equation}
and the CG coefficients $\eta_\sigma$'s are generally real.
The interacting Hamiltonian (\ref{eq:app:h-2}) corresponds to Eq.(\ref{eq:h-int}).

As in Eq.(\ref{eq:app:u-bare}), $U_{\rm bare}$ depends on the field amplitudes $|\Omega_{L}|$ and $|\Omega_{c}|$.
It indicates that the optical phases does not affect the interaction strength.

\section{Effect of the spin imbalance}\label{sec:app:zeeman}

\begin{figure}[t]
	\centering
	\includegraphics[width=0.48\textwidth]{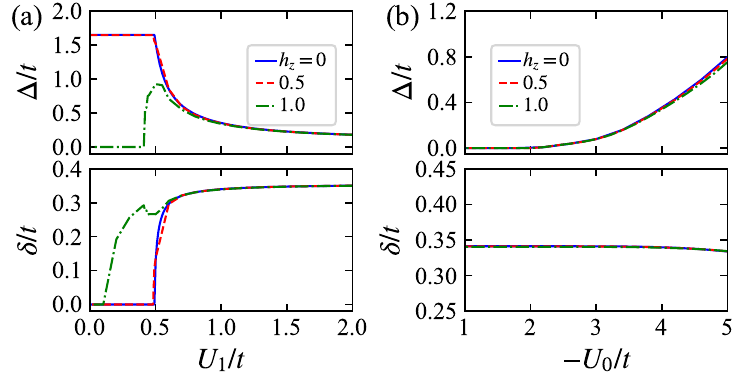}
	\caption{The order parameters $\Delta$ (top) and $\delta$ (bottom) as functions of (a) $U_{1}$ and (b) $U_0$.
		The strength of the Zeeman field $h_z$ is set as 0 (blue-solid line, i.e. the original data from Fig.\ref{fig:order-parameters}), $0.5t$ (red-dashed line), and $1.0t$ (green-dash-dotted line).
		We set $U_0 = -4.0t$ in (a) and $U_1=1.0t$ in (b).}
	\label{fig:order-parameters-zeeman}
\end{figure}

Practically, the spin balance of three spins is an idealized model.
In this section, we investigate the effect of spin imbalance, which is equivalent to an additional Zeeman field,
\begin{equation}
	H_Z = h_z \sum_{j} \Psi_j^\dag \hat{S}_z \Psi_j \,,\label{eq:app:zeeman-hamiltonian}
\end{equation}
where $h_z$ describes the strength of the Zeeman field,
the basis $\Psi_{j}=(\psi_{j,A},\psi_{j,B},\psi_{j,C})^T$ and $\hat{S}_z = {\rm Diag}(1,0,-1)$.
We add the term (\ref{eq:app:zeeman-hamiltonian}) to the BdG Hamiltonian (\ref{eq:h-bdg}),
and repeat the calculations as in Sec.\ref{sec:phase}.
The results are shown in Fig.\ref{fig:order-parameters-zeeman}.
We find the phase boundaries are insensitive to $h_z$ for weak $h_z$,
while the superfluid order vanishes for the superfluid phase in Fig.\ref{fig:order-parameters-zeeman}(a) when $h_z$ is strong.
Furthermore we find that the polarization order parameter $m$ in Eq.(\ref{eq:h-bdg}) is still zero.
This is because the $m$ is defined in the $\hat{S}_x$ direction, thus $h_z$ does not affect it.
The above results indicate the suppression of the superfluid order by mass imbalance, yet the superfluid and SDW orders coexist.

\vfill
\bibliographystyle{apsrev4-1}
\bibliography{ref}
\end{document}